\documentstyle[aps,preprint,prl]{revtex}
\include{epsfig}

\tighten

\begin{document}

\draft


\title{Kinetic Ising model in an oscillating field:
       Finite-size scaling at the dynamic phase transition}

\date{\today}

\author{S.W. Sides,$^{\ast \dagger}$
P.A. Rikvold,$^{\ast \dagger}$
and M.A. Novotny $^{\dagger}$}

\address{$^{\ast}$Center for Materials Research and Technology
         and Department of Physics, \\
$^{\dagger}$Supercomputer Computations Research Institute, \\
Florida State University, Tallahassee, Florida 32306-4130}

\maketitle

\begin{abstract}
We study hysteresis for a two-dimensional,
spin-$1/2$, nearest-neighbor, kinetic Ising
ferromagnet in an oscillating 
field, using Monte Carlo simulations.
The period-averaged magnetization
is the order parameter for a proposed dynamic phase transition (DPT).
To quantify the nature of this transition,
we present the first finite-size scaling study of the DPT
for this model.
Evidence of a diverging correlation length is given,
and we provide estimates of the transition frequency and the
critical indices $\beta$, $\gamma$ and $\nu$.
\end{abstract}


\narrowtext



Although nonequilibrium phase transitions have been studied 
for over two decades, the understanding of their 
universality and scaling properties remains much weaker than for equilibrium 
critical phenomena. An important tool to study 
equilibrium phase transitions is finite-size scaling, 
which has also been applied to 
nonequilibrium transitions in which both the 
driving force and the different states are stationary,
and the nonequilibrium behavior results from the dynamical rules.
Examples include diffusive lattice gas models \cite{leun91,marr96},
Ising models with competing kinetic processes at
different temperatures \cite{aach96}, and 
systems with multiplicative noise \cite{vdbr97}. 

In this letter we present what to our knowledge is the first 
finite-size scaling analysis of dynamic critical phenomena for a system 
in which the nonequilibrium behavior is due to an 
explicit time-dependence of the Hamiltonian, and for which 
the nonequilibrium states are nonstationary in time. 
This is the kinetic Ising model in an oscillating field. 
We focus on
the nature of the dynamic phase transition (DPT) in this
model, which was proposed by several workers 
based on numerical observations 
\cite{lo90,acha95,acha97}.
Our results clarify the nature of the DPT for this specific model and 
also provide evidence for the relevance of 
universality and finite-size scaling 
concepts 
to dynamic phase transitions in non-stationary systems. 

The relaxation of kinetic Ising models 
prepared with all spins aligned in a strong field, which 
is suddenly reversed, models the dynamics of metastable phase
decay in {\it static} field
\cite{rikv94,rich_all}.
Such simulations have been used to study 
magnetization switching in anisotropic ferromagnets. 
The hysteretic response to an {\it oscillating} field
has been studied by, among others, mean-field
\cite{jung90,tome90,luse94} 
and 
Monte Carlo (MC) 
\cite{lo90,acha95,acha97,side_SD,side_all,rao_all,fan95-2}
methods,
and some of the results have been used to
analyze hysteresis loops from experiments on
Fe and Co ultrathin films \cite{he93,jiang95}. 
The relevance of our results thus extends beyond
nonequilibrium statistical 
mechanics to the numerous fields in which kinetic Ising 
systems are used to model specific systems. 

A dynamic phase transition, 
in which the period-averaged magnetization $Q$ passes from a
disordered state ($Q$=$0$) to an ordered state ($Q \neq 0$), 
has been observed in 
mean-field \cite{jung90,tome90,luse94} and 
MC studies
\cite{lo90,acha95,acha97}. 
The location of this transition depends on temperature,
field amplitude, and frequency. 
It can be intuitively understood as a competition
between two time scales: the period of the external field,
$2 \pi / \omega$, and the average lifetime,
$\left < \tau(H_{0}) \right >$,
of the metastable phase following instantaneous field reversal
from $H_{0}$ to $-H_{0}$.
If $2 \pi / \omega < \left < \tau(H_{0}) \right >$
the magnetization cannot fully switch sign
within a single period,
and $Q \neq 0$.
If $2 \pi / \omega > \left < \tau(H_{0}) \right >$
the magnetization follows the field, and $Q$=$0$.

Our previous work on Ising models in sudden field-reversal
simulations has identified distinct decay
regimes in which the decay of the metastable phase proceeds through
nucleation and growth of 
one or more compact droplets of the stable phase.
These different regimes are characterized in great detail in
Refs.~\cite{rikv94,rich_all}.
For this letter, the most important result from past work is
that
the metastable decay mode may change {\it drastically} as the
temperature, field strength, and/or system size are varied.
At weak fields and/or small system size,
the decay proceeds by the nucleation and growth
of a {\it single}
droplet of the stable phase [single-droplet (SD) regime].
For stronger fields and/or larger systems
many droplets contribute to the metastable decay
[multi-droplet (MD) regime].
The MD decay mode can be accurately described
by the classical ``Avrami's law''
for nucleation and growth \cite{seki86}.
For each of these regimes, the statistical properties of the
lifetime of the metastable phase are different, leading to
very different responses in both static and
time-varying \cite{side_SD,side_all} fields.

Here we emphasize two main points.
First, care must be exercised in determining
the temperature and field dependence of the 
DPT boundary for a fixed frequency, because the response of the model
can change qualitatively when the temperature and field strength are
changed.
In fact, our simulations show evidence of a DPT for the MD
regime {\it only}.
Second, if the DPT in the kinetic Ising model is a true critical
phenomenon, then all of the theoretical machinery of
finite-size scaling for equilibrium
phase transitions should be applicable.

The model used is a kinetic, nearest-neighbor
Ising ferromagnet on a
square lattice with periodic boundary conditions
and Hamiltonian
${\cal H }$=$-J \sum_{ {\em \langle ij \rangle}} {\em s_{i}s_{j}}
- H(t) \sum_{i} {\em s_{i}}$.
Here $s_{i}$=$\pm 1$,
$\sum_{ {\em \langle ij \rangle} }$ runs over all
nearest-neighbor pairs, and $\sum_{i}$ runs over all
$L^{2}$ lattice sites.
The ferromagnetic coupling is $J$=$1$.
Each spin is subject to an oscillating field
$H(t)$=$-H_{0} \sin(\omega t)$.
We measure the time-dependent magnetization per site,
$m(t)$=$(1/L^{2}) \sum_{i=1}^{L^{2}} {\em s_{i}(t)}$,
using the Glauber single-spin-flip Monte
Carlo dynamic \cite{bind88}
with updates at randomly chosen sites.
Each attempted spin flip 
from ${\em s_{i}}$ to ${\em -s_{i}}$
is accepted with probability
$W(s_{i} \rightarrow -s_{i})$=$
\exp(- \beta \Delta E_{i}) /[(1 + \exp(- \beta \Delta E_{i}))]$.
Here $\Delta E_{i}$ is the energy change of the system
resulting from an accepted spin flip,
and $\beta = 1/k_{\rm B}T$ where $k_{\rm B}$ is Boltzmann's constant.
The time unit is one Monte Carlo step per spin (MCSS).

All simulations are performed for
three system sizes: $L$=$64$, $90$, and $128$
at $T$=$0.8T_{c}$.
This temperature is sufficiently far below $T_{c}$,
the critical temperature of the Ising model in zero field,
that the thermal correlation length is small compared to
$L$ and the critical droplet radius.
We choose $H_{0}$=$ 0.3J$.
This field amplitude is such that,
for simulations at $0.8T_{c}$ in a static field of $H$=$H_{0}$,
all three system sizes are in the MD regime.
To obtain the raw time-series data, the system was
initially prepared with either a
random arrangement of up and down spins with $m(t=0) \approx 0$,
or with a uniform arrangement with all spins up, $m(t=0)$=$1$.
We recorded $m(t)$ for several values of $\omega$
for approximately $1.7 \times 10^{7}$ MCSS
(for the lowest frequencies, $5.9 \times 10^{5}$ MCSS).
Each of these raw data files
store $t$, $H(t)$, and $m(t)$ in increments of $1$ MCSS
and contain thousands of field periods.
The largest files are about $800$ megabytes and
required $9$ days (one month) to run for $L$=$64$ ($L$=$128$)
on a single node of an IBM sp2. 
This is one of the most extensive MC simulations of hysteresis
in Ising systems to date.

It is useful to think of hysteresis as a competition between two time
scales:
$2 \pi / \omega$ and
$\left < \tau(H_{0}) \right >$.
Therefore we report all our results in terms of the dimensionless
period,
 \begin{equation}
 \label{eq_R}
   R = (2 \pi / \omega) / \left < \tau(H_{0}) \right > \ .
 \end{equation}
The average lifetime has been measured in field reversal
simulations to be
$\left < \tau(H_{0}) \right > \approx 74.5$ MCSS
and is independent of $L$ in the MD regime.

The order parameter of the dynamic phase transition
is the period-averaged magnetization,
 \begin{equation}
 \label{eq_Q}
  Q = \frac{\omega}{2 \pi} \oint m(t) \ dt \ .
 \end{equation}
This  definition removes the field oscillations, so that
$Q$ is a stationary process.
For each frequency
we obtain the probability density of $Q$
by constructing a histogram of the $Q$ values calculated from
each period in the corresponding time series.

\begin{figure}[tbp]
\vspace*{3.5in}
\includegraphics{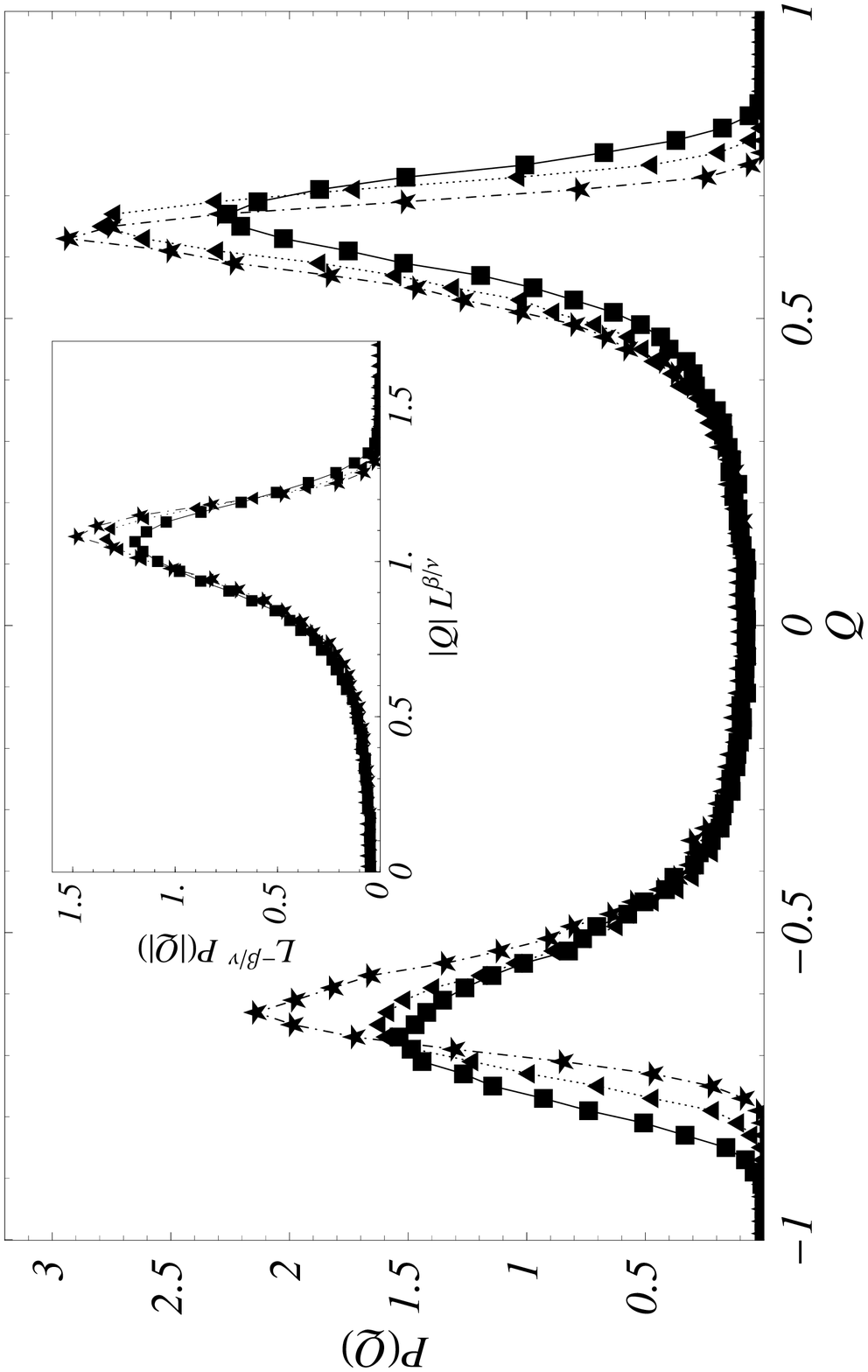}
\caption{\label{fig_unscaled_dist}
Probability densities of the period-averaged magnetization,
$Q$=$(\omega/2 \pi) \oint m(t) dt$ for three system sizes,
$L$=$64$ (squares), $90$ (triangles), and $128$ (stars).
The frequency of the field, $1/R$=$0.2910$ is near the transition.
The asymmetric distributions indicate correlation times
on the order of the simulation time
even for our extremely long runs of $1.7 \times 10^{7}$ MCSS.
Inset: Scaling function
$L^{-\beta / \nu} P(|Q|)$ vs $L^{\beta / \nu} |Q|$.
The value of the scaling exponent is
$(\beta/\nu)_{n=2} \approx 0.111$.
}
\end{figure}

Figure \ref{fig_unscaled_dist} shows the
probability densities of $Q$, $P(Q)$,
for all three system sizes at a frequency near the transition.
(The details of locating the transition frequency are given below.)
Correlation times 
that are significant portions of the total run length
are manifest
in the remaining asymmetry of the distributions.
Estimating the correlation time from the asymmetry in $P(Q)$
we find it to be between $1$ and $10$ percent of the
total simulation length near the transition.
Away from the transition it decreases rapidly.
The behavior is reminiscent of the critical slowing-down
seen in equilibrium simulations, and
even our extensive simulations are not long enough to be fully ergodic.
The asymmetry in the distributions
is {\it not} sensitive to the initial condition of the time series.

\begin{figure}[tbp]
\vspace*{3.5in}
\includegraphics{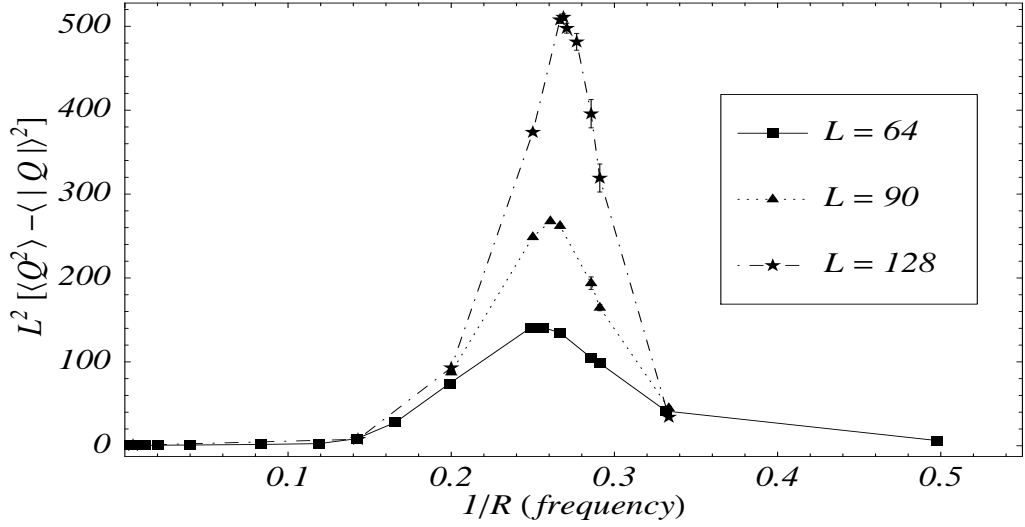}
\caption{\label{fig_suscep}
$L^2 {\rm Var} |Q|$ vs 
dimensionless frequency, $1/R$.
The ``disordered phase,''
($ \left < |Q| \right >$=$0$),
lies on the low-frequency side of the peaks.
The ``ordered phase'',
($ \left < |Q| \right > \neq 0$),
lies on the high-frequency side.
Lines connecting data points are guides to the eye.
The statistical error bars are estimated by partitioning the
data into ten blocks.
Error bars smaller than the symbol sizes are not shown.
}
\end{figure}

At a second-order phase transition there is a divergence in
the susceptibility.
For equilibrium systems, the fluctuation-dissipation theorem
relates the susceptibility to fluctuations in the order parameter.
For the present system, it is not obvious what
the field conjugate to $Q$ might be.
Therefore we cannot measure the susceptibility directly.
However, we can calculate the variance in $|Q|$ as a function of
frequency and study its system size dependence.
We define $X$ as
 \begin{equation}
 \label{eq_suscep}
  X = L^{2} \ {\rm Var}(|Q|)
    = L^{2} \left [ \left < Q^{2} \right > 
      - \left < |Q| \right >^{2} \right ].
 \end{equation}
If the system obeyed a fluctuation-dissipation relation,
$X$ would be proportional to the susceptibility and both
would scale with $L$ in the same manner.
Figure \ref{fig_suscep} shows $X$ vs $1/R$ for three system sizes.
For all three values of $L$, $X$ displays a prominent
peak which increases in height with increasing system size.
This clearly shows finite-size effects in $X$
and implies the existence of a divergent length
associated with the order-parameter correlation function
near the dynamic transition.
The observation that $P(|Q|)$ displays no peak near $|Q|$=$0$ in
the ordered phase
is additional evidence for the second-order (as opposed to first-order)
nature of this transition \cite{eich96}.

\begin{figure}[tbp]
\vspace*{3.5in}
\includegraphics{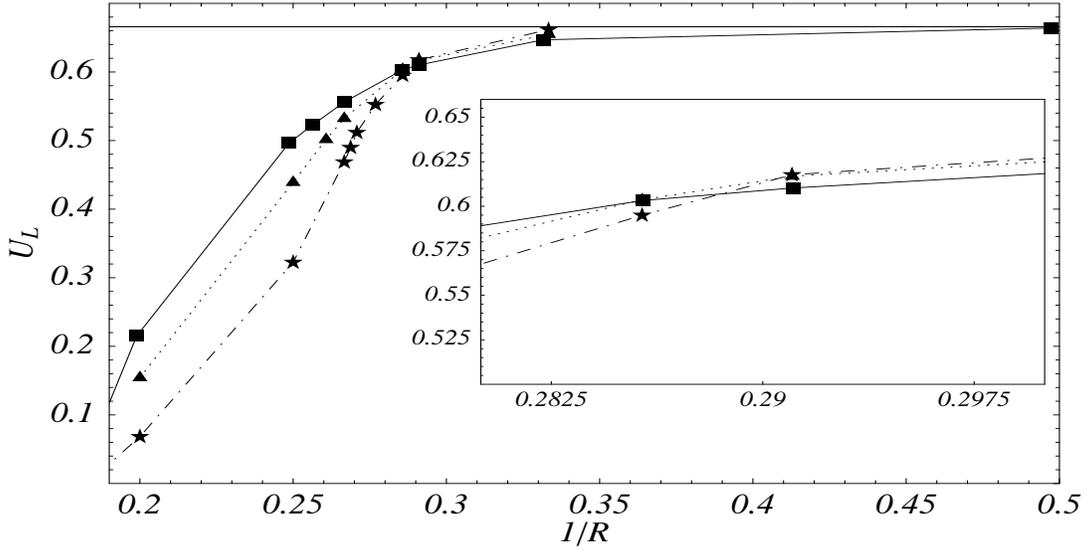}
\caption{\label{fig_cumulant}
Fourth-order cumulant ratio $U_{L}$ vs
dimensionless frequency, $1/R$.
for $L$=$64$, $90$, and $128$.
We use the same system size symbols as in
Fig.~\protect\ref{fig_unscaled_dist}.
The horizontal line marks $U_{L}$=$2/3$.
Lines connecting the data points are guides to the eye.
Inset: area close to the cumulant crossing.
}
\end{figure}

One would like to find the critical exponents associated with
this transition, as well as the frequency at which
the transition occurs.
For the Ising model in zero field, $T_{c}$
is exactly known.
Then, one can use scaling relations
which depend on $T_{c}$
to directly calculate the critical exponents.
In the present case,
no exact solution exists for the transition frequency,
and the scaling relations for $X$ involve
the peak heights and positions.
Both of these are difficult to measure
accurately, even from our extensive data.
The cumulant intersection method \cite{bind88,priv90}
is useful for determining the
location of a second-order transition when the critical exponents
are not known.
We define the ``dynamic'' fourth-order cumulant ratio as
 \begin{equation}
 \label{eq_U}
  U_{L} = 1 - \frac{ \left < |Q|^{4} \right >_{L}}
                   {3 \left < |Q|^{2} \right >_{L}^{2}} \ .
 \end{equation}
where
$\left < |Q|^{n} \right >$=$\int_{0}^{\infty} |Q|^{n} P(|Q|) d|Q|$.
Figure \ref{fig_cumulant} shows $U_{L}$ vs $1/R$.
Above the transition frequency, in the 
$\left < |Q| \right > \neq 0$ phase,
$U_{L}$ approaches $2/3$.
Below the transition frequency, in the
$\left < Q \right >$=$0$ phase,
$U_{L}$ approaches $0$.
At the transition, the cumulant
should have a non-trivial, fixed value, $U^{*}$.
Therefore, the location of the cumulant intersection
gives an estimate of
the transition frequency without foreknowledge of
the critical exponents.
Due to the large spacing of our data and possible
correction-to-scaling
effects, we cannot identify a unique intersection point.
We estimate the location of the intersection by the
crossing of the
two largest system sizes near
$1/R_{c} \approx 0.2910$ with $U_{L}$=$U^{*} \approx 0.61$.

Having estimated the transition frequency, $R_{c}^{-1}$, we
can approximate the critical exponents \cite{bind88,priv90}.
We obtain estimates for
$\beta/\nu$, $\gamma/\nu$, and $1/\nu$
using the two largest system sizes.
At the transition we use scaling relations for
the moments of the order parameter
$(\left < |Q|^{n} \right > \propto L^{-n (\beta / \nu)})$,
the maximum value of the order parameter fluctuations
$(X_{\rm max} \propto L^{\gamma / \nu})$,
and
the position $R_{L}^{-1}$ of the maximum value of the order parameter
fluctuations for a particular system size
($|R_{L}^{-1} - R_{c}^{-1}| \propto L^{-1/ \nu}$).
Using the scaling relations for
either the second or fourth moments of $|Q|$ we estimate
$(\beta / \nu)_{n=2} \approx 0.111$
and $(\beta / \nu)_{n=4} \approx 0.113$.
Our estimates for the other exponents are
$\gamma / \nu \approx 1.84$ and $\nu \approx 1.1$.
Simulations with larger system sizes would be computationally
prohibitive, and smaller system sizes would no longer be in the
MD regime.

The scaled probability distributions of
$|Q|$ are given in the inset of
Fig.~\ref{fig_unscaled_dist} after symmetrizing and
scaling to demonstrate data collapse.
The symmetrizing is equivalent to calculating the distribution
for $|Q|$.
The scaling form is derived in a fashion
analogous to equilibrium
finite-size scaling analysis of
order-parameter distributions \cite{bind88,priv90}.
At the transition, we assume that the mean of the order parameter
scales with $L$ and define the scaling variable
$\tilde{Q}$=$L^{\beta / \nu} |Q|$.
Hence, the scaled probability density for $|Q|$ is given by
 \begin{equation}
 \label{eq_scaling_dist}
  \tilde{P}_{L}(\tilde{Q}) = L^{-\beta / \nu} P(|Q|) \ .
 \end{equation}
The peak positions scale fairly well, the peak heights less so.
This could be due to the following reasons.
The frequency might be sufficiently far
from the transition that single-parameter scaling is not adequate.
There might be corrections to the finite-size scaling
that are large for these relatively small system sizes.
Also, the lack of scaling for the peak heights could
be due to the asymmetry in $P(Q)$ near the transition.
Much longer simulations on larger lattices would be needed to
resolve this issue.

We have also carried out an extensive study of hysteresis in a
kinetic Ising model in the SD regime
where we have found evidence of stochastic resonance,
but {\it no} sign of a dynamic phase transition \cite{side_SD}.
In the introduction we emphasized
that the crossover between the SD and MD
decay regimes depends on temperature, field strength, and system size.
Therefore, the very existence of the dynamic transition, as
well as the details of its critical behavior, may depend sensitively
on all of these parameters.

In conclusion,
we have performed a finite-size scaling study of
a kinetic Ising model in a sinusoidal field in order to clarify
the nature of the dynamic phase transition
conjectured by several authors
\cite{lo90,acha97,side_SD,side_all}.
For this letter, we emphasize that all simulations
were performed in the MD regime.
The behavior of the order-parameter fluctuation, $X$, suggests
a divergent correlation length near the transition frequency.
This behavior motivates the application of finite-size scaling
techniques for second-order phase transitions, analogous to
those used to describe the ferromagnetic/paramagnetic transition
in the Ising model in zero field.
We use the cumulant of the order-parameter distributions to estimate
the value of the transition frequency, $1/R_{c} \approx 0.2910$.
Using scaling relations for the moments and fluctuations
of the order parameter
we estimate the critical exponents to be
$\beta / \nu \approx 0.11$,
$\gamma / \nu \approx 1.84$ and $\nu \approx 1.1$.
Our results for $\beta / \nu$
and $\gamma / \nu$ are close to the
two-dimensional Ising values for the analogous exponent ratios.
The result,
$2(\beta / \nu) + (\gamma / \nu) \approx 2.06 \approx d$,
gives a tantalizing indication that hyperscaling may be obeyed.
Also, our value for the cumulant intersection,
$U^{*} \approx 0.61$, agrees
with an extremely precise transfer
matrix calculation of $U^{*}$=$0.6106901(5)$
for the two-dimensional Ising model \cite{blot93}.
To precisely calculate the critical frequency,
critical exponents, and determine
the universality class of the transition would require simulations
for more frequencies, larger systems, and for immensely longer
run times to improve the statistics.
More detailed analysis of this problem
could resolve specific questions about the DPT for
the Ising model, such as the existence of a tricritical
point in the dynamic phase diagram, in addition
to elucidating general questions
concerning the nature of nonequilibrium phase transitions
in models with explicitly time-dependent Hamiltonians.

We would like to thank
G. Brown,
W. Janke,
W. Klein,
M. Kolesik,
G. Korniss,
M. Acharyya,
and
J. Vi\~{n}als
for useful discussions.
Supported in part by
the FSU Center for Materials Research and Technology (MARTECH),
by the
FSU Supercomputer Computations Research Institute (SCRI) under DOE
Contract No.\ DE-FC05-85ER25000,
and by NSF
Grants No.\ DMR-9315969, DMR-9634873, and DMR-9520325.


\newpage

\newpage

\end{document}